\documentclass[conference]{IEEEtran}
\usepackage[pdftex]{graphicx}
\usepackage{amsmath}
\usepackage{amssymb}
\graphicspath{{figures/}}

\begin{document}
\title{Robust RF Data Normalization for Deep Learning}

\author{\
\IEEEauthorblockN{Mostafa~Sharifzadeh, Habib~Benali, and Hassan~Rivaz}
\IEEEauthorblockA{Department of Electrical and Computer Engineering\\Concordia University\\ Montreal, QC, Canada\\mostafa.sharifzadeh@mail.concordia.ca}}
\maketitle

\begin{abstract}
Radio frequency (RF) data contain richer information compared to other data types, such as envelope or B-mode, and employing RF data for training deep neural networks has attracted growing interest in ultrasound image processing. However, RF data is highly fluctuating and additionally has a high dynamic range. Most previous studies in the literature have relied on conventional data normalization, which has been adopted within the computer vision community. We demonstrate the inadequacy of those techniques for normalizing RF data and propose that individual standardization of each image substantially enhances the performance of deep neural networks by utilizing the data more efficiently. We compare conventional and proposed normalizations in a phase aberration correction task and illustrate how the former enhances the generality of trained models.
\end{abstract}

\IEEEpeerreviewmaketitle

\section{Introduction}
In the field of processing medical ultrasound images, deep learning-based methods have gained considerable interest in recent years due to their superior performance compared to traditional approaches across various tasks. These techniques leverage the power of neural networks to enhance image quality and aid in diagnostics and have found application in tasks such as beamforming \cite{Lu2022, Goudarzi2022, Lei2023}, phase aberration correction \cite{Sharifzadeh2020, Koike2023}, speckle reduction \cite{Asgariandehkordi2023}, image segmentation \cite{Sharifzadeh2022}, image registration \cite{Salari2023}, elastography \cite{Tehrani2023}, quantitative ultrasound \cite{KafaeiZadTehrani2023, Soylu2023,Soylu2023a}, and more.

Recently, there has been an increasing adoption of radio frequency (RF) data in deep learning-based approaches due to the fact that these methods prove more effective with more data, and RF data inherently contains richer information compared to envelope or B-mode data. RF data has a higher fidelity stemming from its raw and unprocessed form and contains complex details about the interaction between ultrasound waves and tissue structures. This makes it particularly well-suited for deep learning techniques, where these methods can leverage the complexity of RF data to detect subtle tissue differences, texture variations, and acoustic properties that might be missed in envelope or B-mode data. However, the highly fluctuating nature of RF data poses a challenge for neural networks to learn effectively during training, given that regions exhibiting comparable patterns might not appear very similar in RF data representation from a network's perspective. This challenge worsens in the presence of bright specular reflectors. These large echoes arise due to the substantial differences in amplitudes of raw ultrasound signals, caused by variations in tissue density, acoustic impedance, and other factors, resulting in a high dynamic range (HDR) of signal amplitudes.

The RF data may be acquired under varying power settings and from diverse sources, such as various simulation packages and ultrasound machines, and needs to be normalized. While numerous researchers opt for the utilization of min-max scaling to align the RF data within a predefined range, such as [-1, 1], or less conventional ranges like [0, 1], it is important to recognize the susceptibility of these techniques to HDR RF data. Although min-max scaling exhibit efficacy in the context of natural images, it may not be as effective for fluctuating RF signals with very high HDR.

In this study, we demonstrate the inadequacy of the conventional min-max scaling techniques for normalizing RF data and illustrate how large amplitudes generated by a typical structure, such as a bright specular reflector, introduce challenges to the learning process of a neural network by preventing it from utilizing the data effectively. Additionally, we propose that employing a robust normalization method substantially improves the network's performance.

\section{Methodology}
\subsection{Robust Normalization}
Let us denote the RF data of the ultrasound image by $RF(x,y)$. A conventional normalization technique involves dividing the RF data by its maximum absolute value, resulting in $RF_{MaxAbs}(x,y)$ as given by
\begin{equation}
	\label{eq-1}
	\begin{aligned}
		RF_{MaxAbs}(x,y)=\frac{RF(x,y)}{\max\left|RF(x,y)\right|}.
	\end{aligned}
\end{equation}
This step plays a critical role in transforming the RF data, acquired from various simulation packages and ultrasound machines, to a consistent range of [-1, 1]. We propose that following the prior step, applying individual standardization to the image substantially enhances the performance of deep neural networks by utilizing the data more efficiently:
\begin{equation}
	\label{eq-2}
	\begin{aligned}
		RF_{Robust}(x,y)=\frac{RF_{MaxAbs}(x,y)}{\sigma}
	\end{aligned}
\end{equation}
where $\sigma$ is the standard deviation of values across $RF_{MaxAbs}(x,y)$.

By individually dividing each image by its corresponding standard deviation, the RF data is efficiently normalized with regard to its variability, which mitigates the impact of large echoes in the process of comparing different images. It is worth noting that this approach yields a distinct outcome compared to the well-known standardization technique applied within deep learning frameworks, which relies on dataset-wide statistics.
In this particular context, the term ``robust’' indicates that the RF data from regions with similar patterns undergo a transformation that results in an increased similarity between the scale of their amplitude values. This interpretation of ``robust’' should not be confused with the notion that the standard deviation value is insensitive to the larger amplitude of bright echoes.

\subsection{Dataset}
We synthesized 100 single plane-wave images using a full synthetic aperture scan, each representing a randomly aberrated version of an identical phantom measuring 45 mm laterally and 40 mm axially containing two anechoic cysts. The full synthetic aperture scan was simulated using Field II \cite{Jensen1996}, and images were synthesized as elaborated in \cite{Sharifzadeh2023}.
Additionally, among the aberrated images, one of them was selected and subsequently replicated. However, for the replicated version, we added a point target into the phantom before running simulations, introducing bright echoes into the RF data. Furthermore, we created non-aberrated versions of both phantoms, with and without the point target, for visualization purposes.
Transducer settings used for simulation were similar to those of the 128-element linear array L11-5v (Verasonics, Kirkland, WA). The central and sampling frequencies were set to 5.208 MHz and 20.832 MHz, respectively. To ensure accurate numerical results in Field II simulations, we initially set the sampling frequency to 104.16 MHz and then downscaled the simulated data by a factor of 5.

\subsection{Phase Aberration Correction Task}
We evaluated the effectiveness of the robust RF data normalization in a phase aberration correction task by conducting a simple yet enlightening experiment. To correct the phase aberration effect, an aberration to aberration approach \cite{Sharifzadeh2023} was employed. In this approach, the network maps distinct randomly aberrated versions of the same realization to each other during the training phase and is expected to output a corrected version in the inference phase. Interested readers may refer to \cite{Sharifzadeh2023} for more details.
Out of the 100 aberrated versions of the phantom without the point target, 99 versions served as a training set, and during each epoch, each of the 99 versions was randomly mapped to another one. The remaining version, along with its replica containing a point target, was reserved for evaluation purposes.

Two U-Nets \cite{Ronneberger2015} were trained using identical settings on the same training set with no point target. The only difference lay in the preprocessing steps: for the first network, each image in the dataset was normalized by a division with its maximum value, whereas for the second network, a robust normalization approach was employed. In the inference phase, the test images were also subjected to normalization, consistent with the method employed during the network's training.

The networks were trained for 1000 epochs, with a sigmoid function as the activation function of the last layer, and the batch size was 32. In both cases, the dataset was standardized by subtracting the mean and dividing it by its standard deviation. We utilized an adaptive mixed loss \cite{Sharifzadeh2023} as the loss function and Adam \cite{Kingma2015} with a zero weight decay as the optimizer. The learning rate was initially set to $10^{-3}$ and halved at epoch 500.

\section{Results and Discussion}
\begin{figure}
	\centering
	\includegraphics[width=0.9999\linewidth]{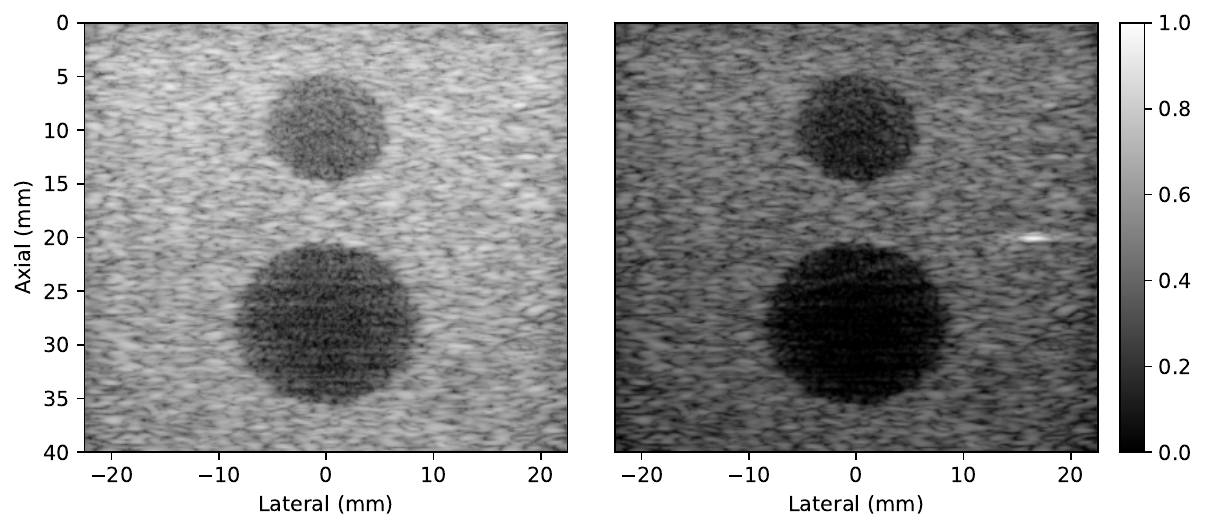}
	\caption{A pair of ultrasound images simulated to be exactly identical using the Field II simulation package, where the only distinction between them lay in the presence of a point target within the second one. Both images are normalized in the same range and shown on the same dynamic range.}
	\label{fig-bmodes}
\end{figure}

\begin{figure}
	\centering
	\includegraphics[width=0.9999\linewidth]{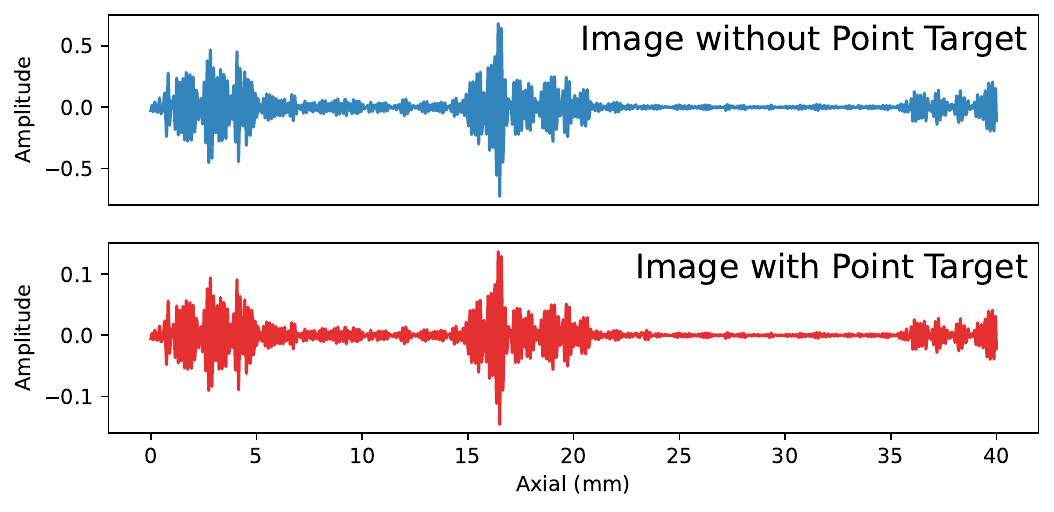}
	\caption{The RF data associated with the middle column of images with and without the point target shown in Fig. \ref{fig-bmodes}. The RF data were normalized by dividing them by their maximum absolute values across the entire image. In the top signal, the range of amplitude is roughly 5 times larger.}
	\label{fig-rf_middle_col}
\end{figure}

\begin{figure}
	\centering
	\includegraphics[width=0.9999\linewidth]{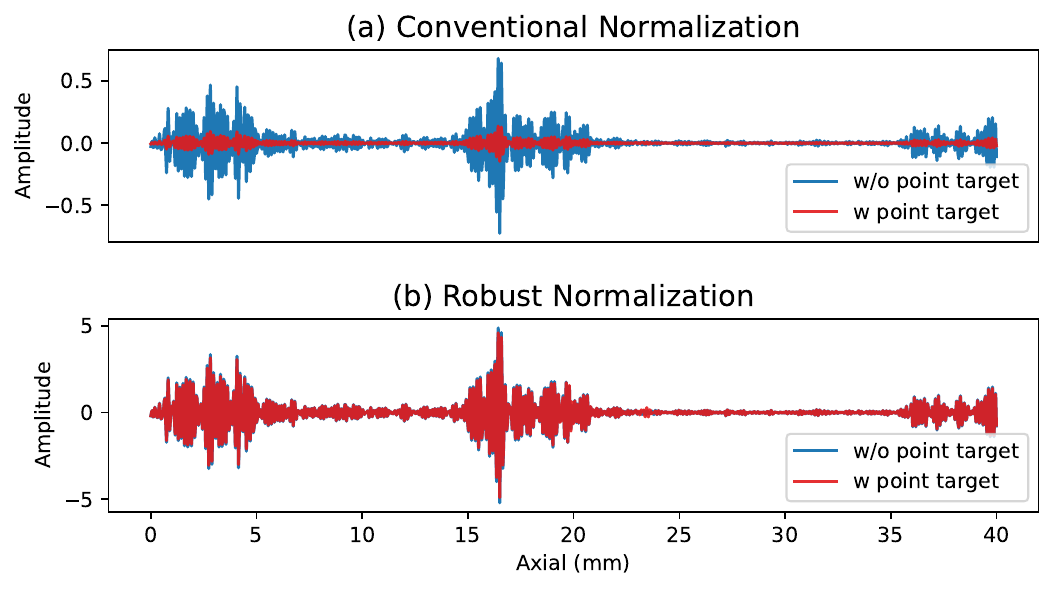}
	\caption{The RF data associated with the middle column of images with and without the point target shown in Fig. \ref{fig-bmodes}. In the bottom figure, the two signals almost overlap.}
	\label{fig-rf_comparison}
\end{figure}

\begin{figure}
	\centering
	\includegraphics[width=0.9999\linewidth]{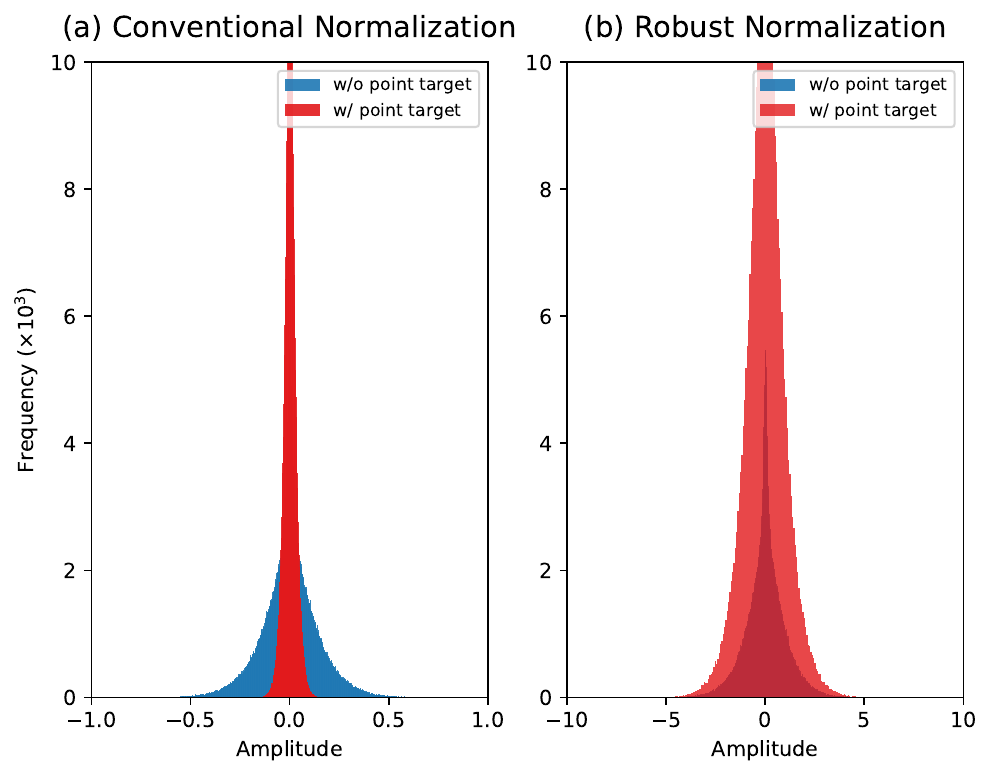}
	\caption{Histograms of RF data associated with the images shown in Fig. \ref{fig-bmodes}, where the data was normalized by (a) the conventional method and (b) the robust method.}
	\label{fig-histogram_comparison}
\end{figure}

\begin{figure*}
	\centering
	\includegraphics[width=0.9999\linewidth]{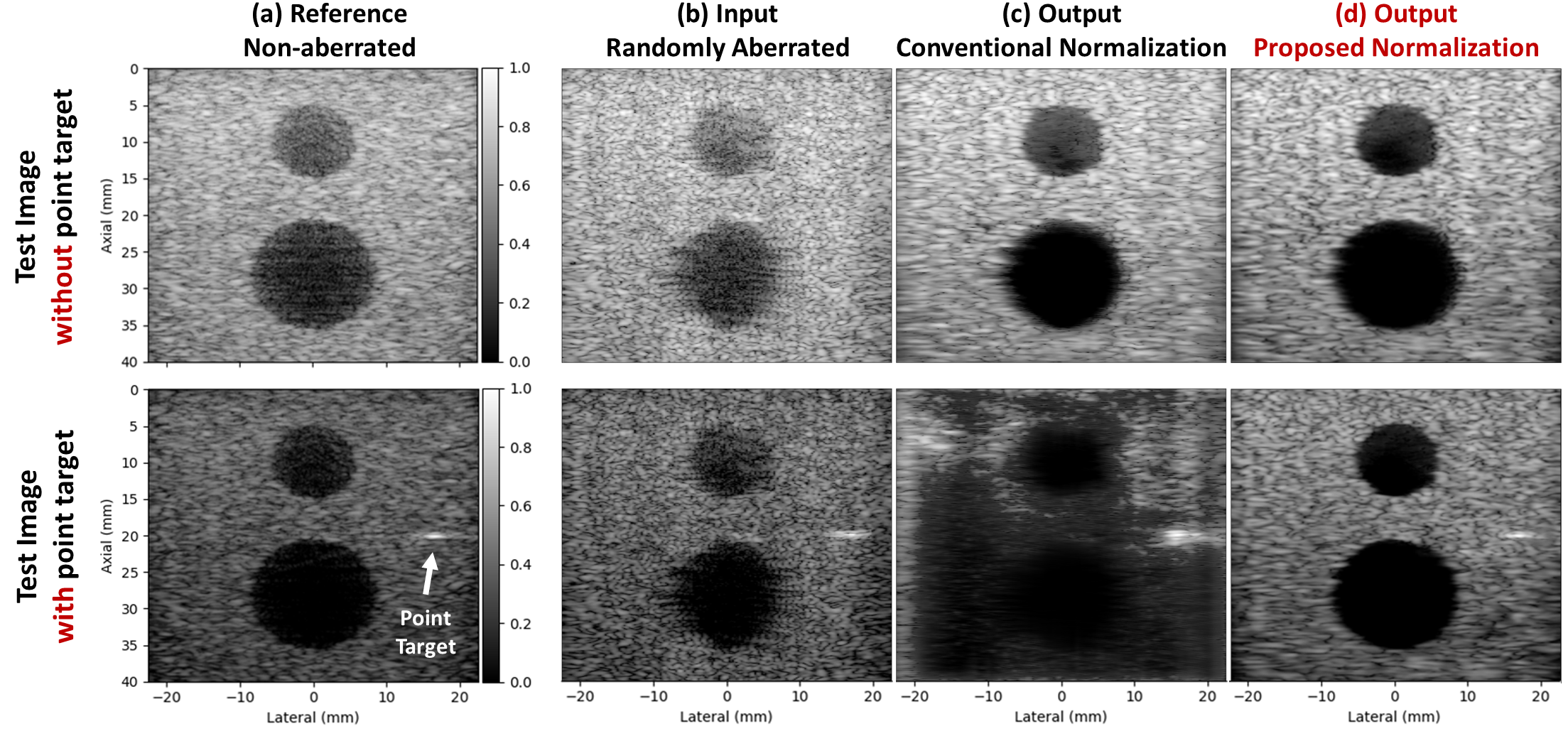}
	\caption{Evaluating the efficacy of the robust normalization technique in a phase aberration correction task. (a) Non-aberrated reference images with and without point targets, reconstructed using DAS. (b) Randomly aberrated inputs with and without point target. (c) Output from the network trained on conventionally normalized data, utilizing similarly normalized inputs. (d) Output from the network trained on robustly normalized data, also with inputs normalized in a similar manner. All images are displayed in the same scale on a 50 dB dynamic range.}
	\label{fig-results}
\end{figure*}

Consider a pair of ultrasound images meticulously simulated to be identical using the Field II simulation package. The only distinction between these images lay in the presence of a point target exclusively within the second one. The RF data of each simulated image was normalized by dividing it by its maximum absolute value. The B-mode images are shown in Fig. \ref{fig-bmodes}. To isolate specific structures and analyze their amplitude variations, the RF data corresponding to the middle column of each image is plotted in Fig. \ref{fig-rf_middle_col}. It is evident that while the fundamental pattern of the RF data remains consistent, there is a substantial difference between their amplitude ranges. As expected, the image containing the point target exhibits markedly lower amplitudes compared to the image lacking the said point target. The RF data for both images are superimposed in Fig. \ref{fig-rf_comparison}(a), where the amplitude corresponding to the image with the point target became almost negligible compared to the other image. Therefore, from a network's perspective, an identical pattern in one image could be construed as a constant signal with a zero amplitude (e.g., an anechoic region) in the other image. Consequently, the network is unable to transfer its learned knowledge from one signal to the other. 

By employing the proposed robust normalization to preprocess the RF data of images, a noticeable improvement was observed in addressing the previously mentioned issue. As illustrated in Fig. \ref{fig-rf_comparison}(b), the amplitudes of RF signals, which were initially expected to be identical, subsequently fell within the same range.
Note that in the context of conventional normalization, the RF data was normalized by dividing each image by its maximum absolute value present within the whole image. Consequently, as shown in Fig. \ref{fig-rf_comparison}, the expected outcome was that the amplitude of the RF signal would be confined within the range of [-1, 1]. However, by applying the robust normalization, the amplitude values expanded to cover a broader range beyond [-1, 1]. This expansion in range does not raise any concerns as long as the same robust normalization approach is consistently applied during both the training and testing phases. The histograms shown in Fig. \ref{fig-histogram_comparison} represent the full RF data for both images, which were normalized using conventional and robust techniques.

To evaluate the effectiveness of the proposed robust normalization technique in a phase aberration correction task, we trained two different networks utilizing an identical dataset devoid of any point targets. Therefore, neither of the networks had been exposed to any point targets during the training phase. Nonetheless, during the training process, the initial network's dataset images underwent normalization through the conventional method, while the second network's images were normalized using the robust approach.
We provided both networks with two aberrated test images: one containing a point target, and the other without, as illustrated in Fig. \ref{fig-results}(b), with all images displayed in the same scale on a 50 dB dynamic range.
As we can see in the outputs shown in (c), the inclusion of a point target within the RF data introduced large echoes, and applying a conventional normalization resulted in the compression of all other values towards proximity to zero, producing an inferior output. In contrast, as shown in (d), applying robust normalization effectively mitigated the impact of those large echoes on the remaining data. It enabled the network to leverage its learned knowledge from the other image and output a corrected image, even in the case where its training set did not include any point targets.
It is essential to recognize that adding bright specular reflectors to the training dataset can indeed help the network in enhancing its ability to handle these features. However, it is crucial to emphasize that such augmentation does not serve as a replacement for the robust normalization technique. Even by adding those reflectors with different random intensities to the dataset, the network still may fail to establish a coherent relationship between similar structures when their amplitudes occur on different scales. This effect is comparable to dividing the dataset into multiple smaller subsets based on the presence of large echoes and their amplitudes, which subsequently results in a suboptimal efficiency.

\section{Conclusion}
We investigated the importance of normalizing RF data on the performance of deep learning-based approaches and demonstrated the inadequacy of conventional min-max scaling techniques, particularly in a phase aberration correction task. We showed that standardizing RF data individually, considering the variability within each image, leads to a more consistent range of amplitude values for similar regions across different images. This process alleviates the impact of large echoes and helps the network seamlessly transfer its learned knowledge across images, resulting in higher performance.

\section*{Acknowledgment}
The authors would like to thank Natural Sciences and Engineering Research Council of Canada (NSERC) for funding.

\bibliographystyle{IEEEtran}
\bibliography{bibliography}

\end{document}